\documentclass[Physsubmission, Phys]{SciPost}

\hypersetup{
    colorlinks,
    linkcolor={red!50!black},
    citecolor={blue!50!black},
    urlcolor={blue!80!black}
}

\usepackage[bitstream-charter]{mathdesign}
  \usepackage{xspace}
 \usepackage{amsfonts}
 \usepackage{verbatim}
\usepackage{tabularx}

\newcolumntype{C}[1]{>{\centering\arraybackslash}p{#1}}

\newcommand{\loqcd}{\rm LO_{\rm QCD}\xspace}
\newcommand{\loew}{\rm LO_{\rm EW}\xspace}
\newcommand{\nloone}{{\rm NLO}_1\xspace}
\newcommand{\nlotwo}{{\rm NLO}_2\xspace}
\newcommand{\nlothree}{{\rm NLO}_3\xspace}

\def\reffi#1{\mbox{Fig.~\ref{#1}}}
\def\reffis#1{\mbox{Figs.~\ref{#1}}}

\def\citere#1{\mbox{Ref.~\cite{#1}}}

\newcommand{\newc}{\newcommand}
\newc{\beq}{\begin{equation}}
\newc{\eeq}{\end{equation}}
\newc{\bit}{\begin{itemize}}
\newc{\eit}{\end{itemize}}
\newc{\ben}{\begin{enumerate}}
\newc{\een}{\end{enumerate}}
\newc{\bce}{\begin{center}}
\newc{\ece}{\end{center}}
\newc{\bfi}{\begin{figure}}
\newc{\efi}{\end{figure}}

\newcommand{\rT}{{\mathrm{T}}}

\newcommand{\eg}{\emph{e.g.}\ }

\newcommand{\GeV}{\ensuremath{\,\text{GeV}}\xspace}

\newcommand{\Pp}{\ensuremath{\text{p}}}
\newcommand{\Pe}{\ensuremath{\text{e}}\xspace}
\newcommand{\Pb}{\ensuremath{\text{b}}\xspace}

\newcommand{\Pt}{\ensuremath{\text{t}}\xspace}

\newcommand{\PW}{\ensuremath{\text{W}}\xspace}

\newcommand{\Pl}{\ell}
                                    
\newcommand{\Mt}{\ensuremath{m_\Pt}\xspace}

\newcommand{\recola}{{\sc Recola}\xspace}

\newcommand{\mocanlo}{{\sc MoCaNLO}\xspace}
\newcommand{\collier}{{\sc Collier}\xspace}

\newcolumntype{.}{D{.}{.}{-1}}
\newcolumntype{d}[1]{D{.}{.}{#1}}
\colorlet{tableoverheadcolor}{gray!37.5}
\colorlet{tableheadcolor}{gray!25}
\colorlet{tablerowcolor}{gray!12.5}

\def\draftdate{\relax}
\def\mda{\relax}
\def\mua{\relax}
\def\mla{\relax}
\def\draft{
\def\thtystars{******************************}
\def\sixtystars{\thtystars\thtystars}
\typeout{}
\typeout{\sixtystars**}
\typeout{* Draft mode!
         For final version remove \protect\draft\space in source file *}
\typeout{\sixtystars**}
\typeout{}
\def\draftdate{\today}
\def\mua{\marginpar[\boldmath\hfil$\uparrow$]%
                   {\boldmath$\uparrow$\hfil}\color{black}%
                    \typeout{marginpar: $\uparrow$}\ignorespaces}
\def\mda{\color{red}\marginpar[\boldmath\hfil$\downarrow$]%
                   {\boldmath$\downarrow$\hfil}%
                    \typeout{marginpar: $\downarrow$}\ignorespaces}
\def\mla{\marginpar[\boldmath\hfil$\rightarrow$]%
                   {\boldmath$\leftarrow $\hfil}%
                    \typeout{marginpar: $\leftrightarrow$}\ignorespaces}
\def\Mua{\marginpar[\boldmath\hfil$\Uparrow$]%
                   {\boldmath$\Uparrow$\hfil}\color{black}%
                    \typeout{marginpar: $\uparrow$}\ignorespaces}
\def\Mda{\color{red}\marginpar[\boldmath\hfil$\Downarrow$]%
                   {\boldmath$\Downarrow$\hfil}%
                    \typeout{marginpar: $\downarrow$}\ignorespaces}
\def\Mla{\marginpar[\boldmath\hfil\textcolor{red}{$\Rightarrow$}]%
                   {\boldmath\textcolor{red}{$\Leftarrow $}\hfil}%
                    \typeout{marginpar: $\leftrightarrow$}\ignorespaces}
\overfullrule 5pt
\oddsidemargin 15mm
\marginparwidth 29mm
}

\newcommand{\mc}{\mathcal}

\let\as\alphas
\newcommand{\pt}[1]{p_{\rT,{#1}}}

\begin{document}

\begin{center}{\Large \textbf{
NLO electroweak and QCD corrections to off-shell ttW production at the LHC\\
}}\end{center}

\begin{center}
A. Denner\textsuperscript{1} and
G. Pelliccioli\textsuperscript{1$\star$}
\end{center}

\begin{center}
{\bf 1}
University of W\"urzburg,
  Instit\"ut f\"ur Theoretische Physik und Astrophysik,
  Emil-Hilb-Weg 22, 97074 W\"urzburg (Germany)\\
* giovanni.pelliccioli@physik.uni-wuerzburg.de
\end{center}

\begin{center}
\today
\end{center}


\definecolor{palegray}{gray}{0.95}
\begin{center}
\colorbox{palegray}{
  \begin{tabular}{rr}
  \begin{minipage}{0.1\textwidth}
    \includegraphics[width=35mm]{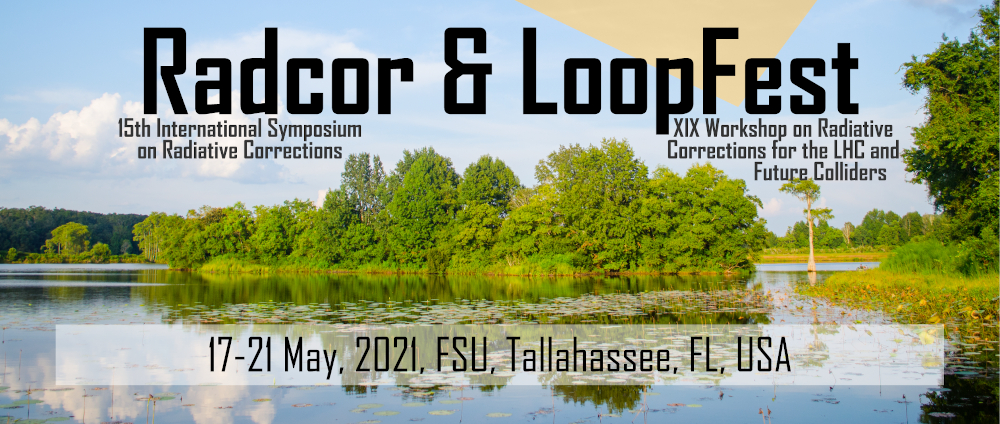}
  \end{minipage}
  &
  \begin{minipage}{0.85\textwidth}
    \begin{center}
    {\it 15th International Symposium on Radiative Corrections: \\Applications of Quantum Field Theory to Phenomenology,}\\
    {\it FSU, Tallahasse, FL, USA, 17-21 May 2021} \\
    \doi{10.21468/SciPostPhysProc.?}\\
    \end{center}
  \end{minipage}
\end{tabular}
}
\end{center}

\section*{Abstract}
{\bf
  The foreseen luminosities for the next LHC runs will enable
precise differential measurements of the associated production
of top--antitop pairs with a W boson. Therefore, providing
accurate theory predictions for this process is needed for
realistic final states.
We present the first combination of NLO electroweak and
QCD corrections to off-shell ttW$^+$ production in the
three-charged-lepton channel, including non-resonant effects,
full spin-correlations and interferences.
Such radiative corrections comprise the electroweak and QCD
corrections to the leading QCD order, and the QCD corrections
to the leading electroweak order.
}

\section{Introduction}
\label{sec:intro}
The associated production of top--antitop pairs with $\PW$~bosons
represents one of the heaviest signatures at the LHC and an important
process to study, both as a probe
of the Standard Model (SM) and as a window to new-physics effects. It gives direct access to the
coupling of top~quarks to electroweak (EW) bosons in or beyond the SM \cite{Dror:2015nkp,Buckley:2015lku,Bylund:2016phk} and
it is expected to enhance the sensitivity to the $\Pt\overline{\Pt}$ charge asymmetry
\cite{Maltoni:2014zpa}. This process is also a relevant background to the associated $\Pt\overline{\Pt}$
production with a Higgs boson \cite{Maltoni:2015ena}. Recent ATLAS and CMS public results \cite{Sirunyan:2017uzs,Aaboud:2019njj} and
improved analyses \cite{ATLAS-CONF-2019-045,CMS-PAS-HIG-17-004} for $\Pt\overline{\Pt}\PW$
point in the direction of a tension between data and SM predictions, which has not been addressed yet in spite of strong efforts in the
theoretical community. Since the high-luminosity run of the LHC will allow for differential
measurements in $\Pt\overline{\Pt}\PW$, it is essential to provide precise SM predictions for this process in specific
decay channels.
The next-to-leading-order (NLO) QCD and EW corrections are known since several years for the inclusive production 
\cite{Maltoni:2014zpa,Maltoni:2015ena,Frixione:2015zaa,Frederix:2017wme,Frederix:2018nkq}.
Soft-gluon resummation \cite{Li:2014ula,Broggio:2016zgg,Kulesza:2018tqz,Broggio:2019ewu,Kulesza:2020nfh} and multi-jet merging
\cite{vonBuddenbrock:2020ter,Frederix:2021agh} have been performed for inclusive production.
The decay modelling has been tackled in the narrow-width approximation (NWA)
at NLO QCD \cite{Campbell:2012dh} and with matching to parton shower \cite{Garzelli:2012bn}.
The subleading NLO QCD corrections to the LO EW have been computed in the NWA, including
spin correlations and parton-shower effects \cite{Frederix:2020jzp,1843174}.
The first predictions for off-shell $\Pt\bar{\Pt}\PW$ production in
the three-charged-lepton channel
have appeared recently \cite{Bevilacqua:2020pzy,Denner:2020hgg,Bevilacqua:2020srb}
and have been also compared to NWA results matched with parton-shower \cite{Bevilacqua:2021tzp}.
Based on Ref.~\cite{Denner:2021hqi}, we present the first complete fixed-order description
of the off-shell production of $\Pt\bar{\Pt}\PW^+$ in the three-charged-lepton decay channel,
combining NLO QCD and EW corrections which are sizeable at the LHC@13TeV.

\section{Description of the calculation}
\label{sec:descr}

\begin{figure*} 
  \centering
  \includegraphics[scale=0.6]{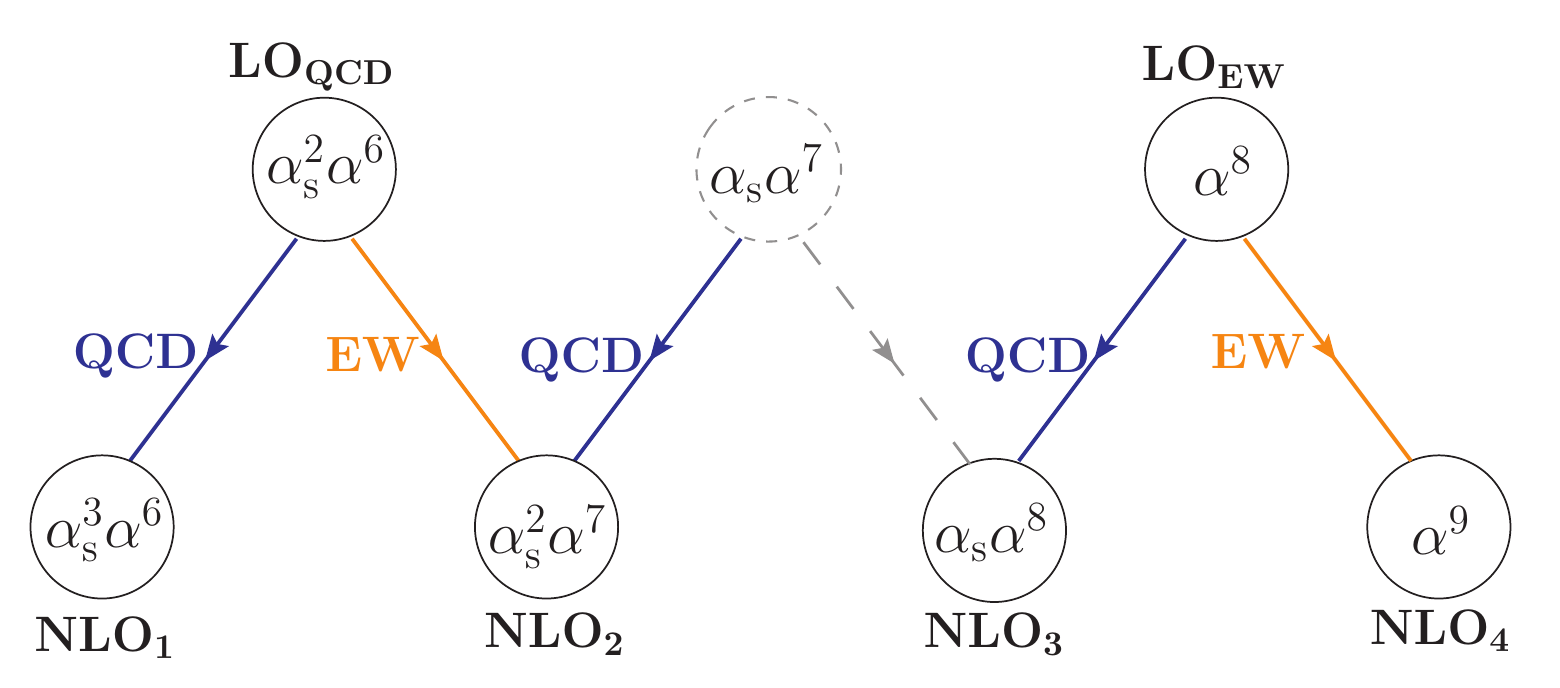}  
  \caption{
    Perturbative orders contributing at LO and NLO
    to $\Pt\overline{\Pt}\PW$ production in the three-charged-lepton decay channel.
  }\label{fig:orders}
\end{figure*}

\begin{figure*} 
  \centering
  \includegraphics[scale=0.26]{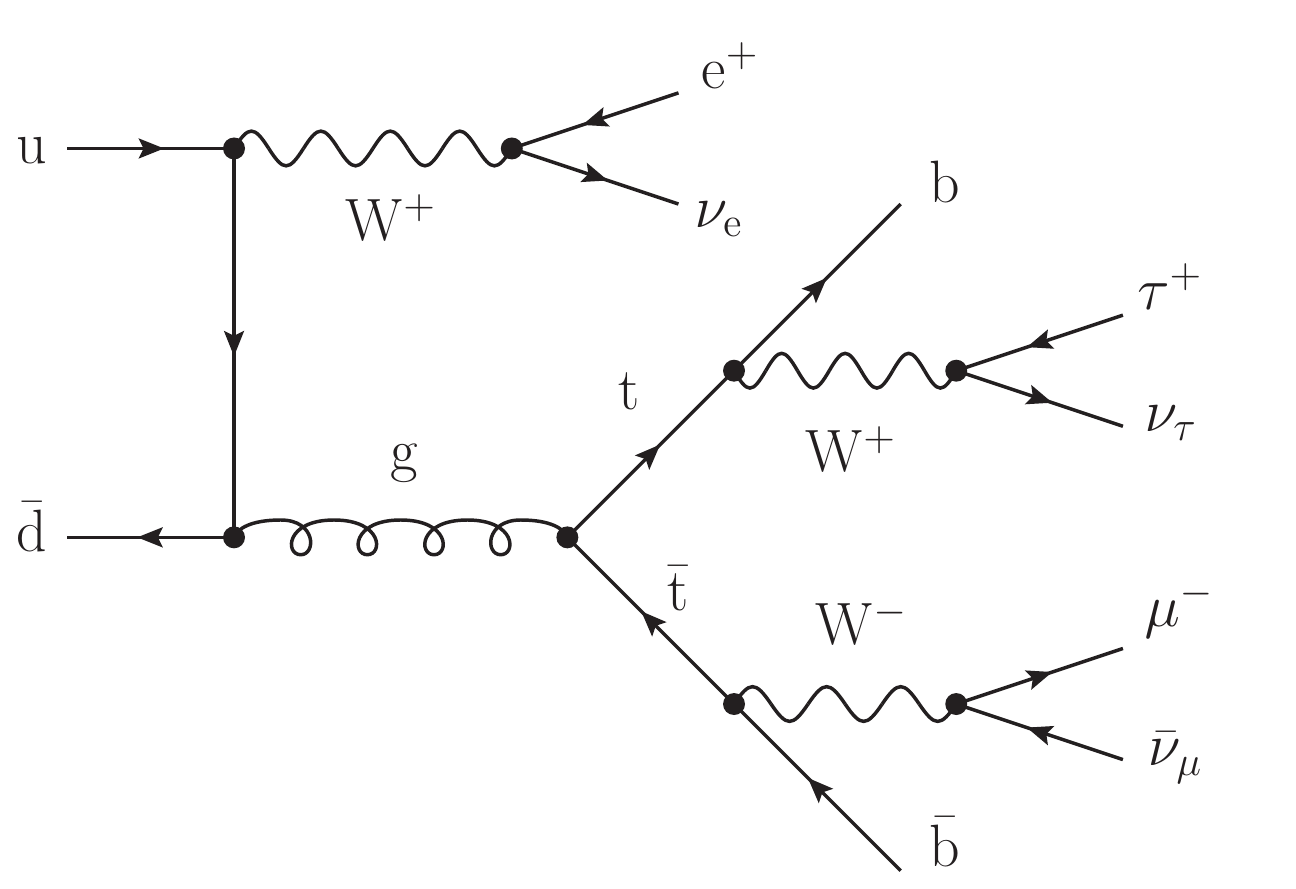}
  \includegraphics[scale=0.26]{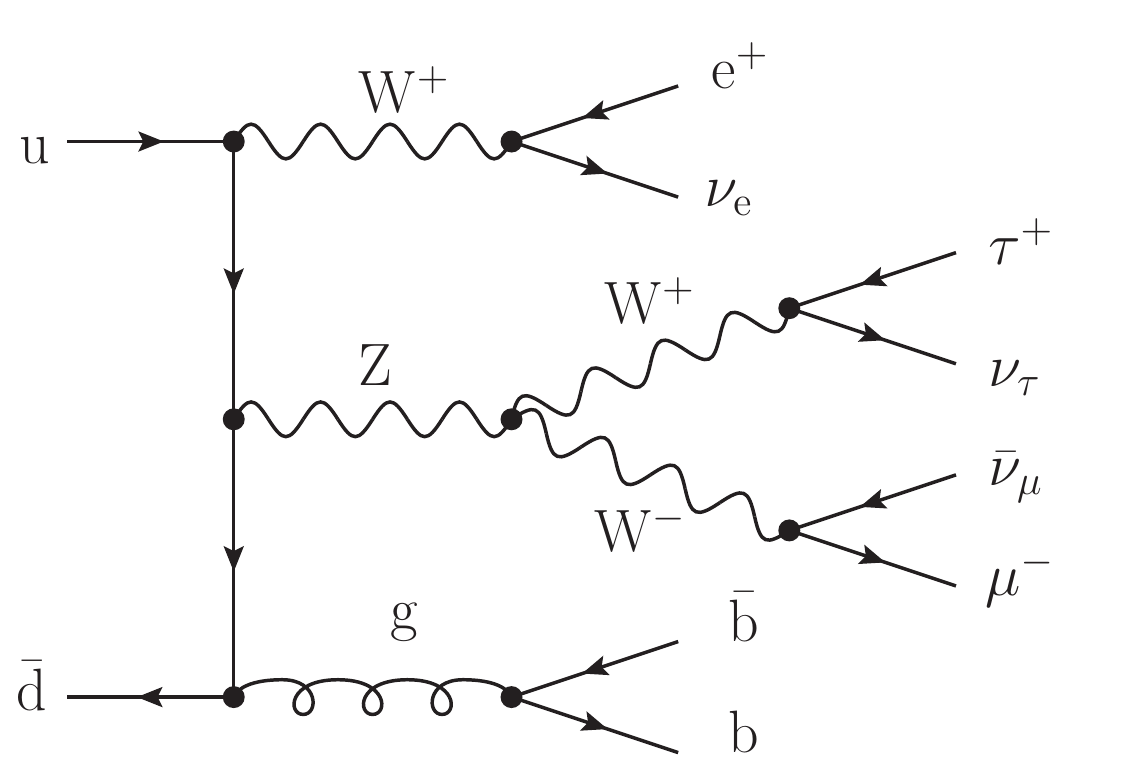}
  \includegraphics[scale=0.26]{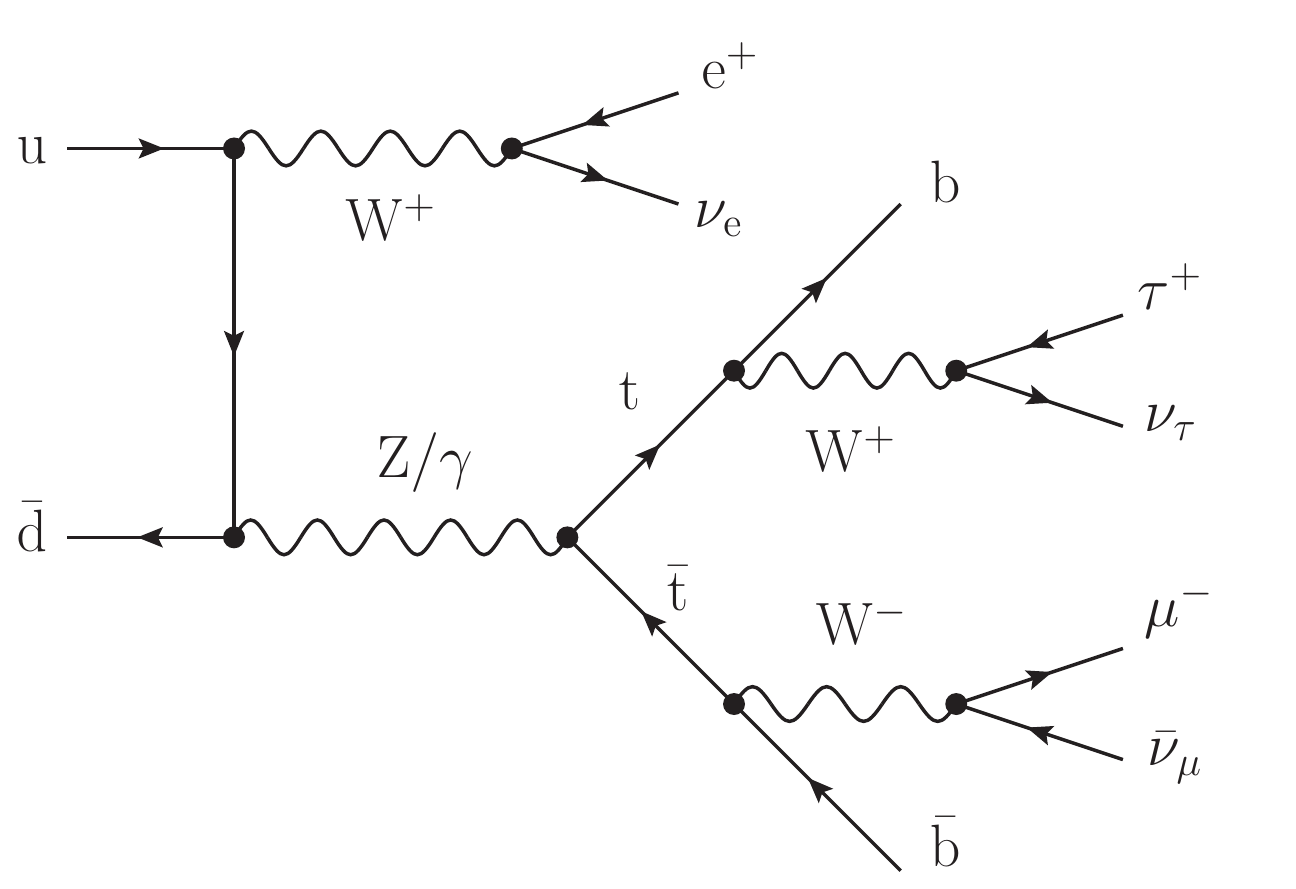}
  \includegraphics[scale=0.26]{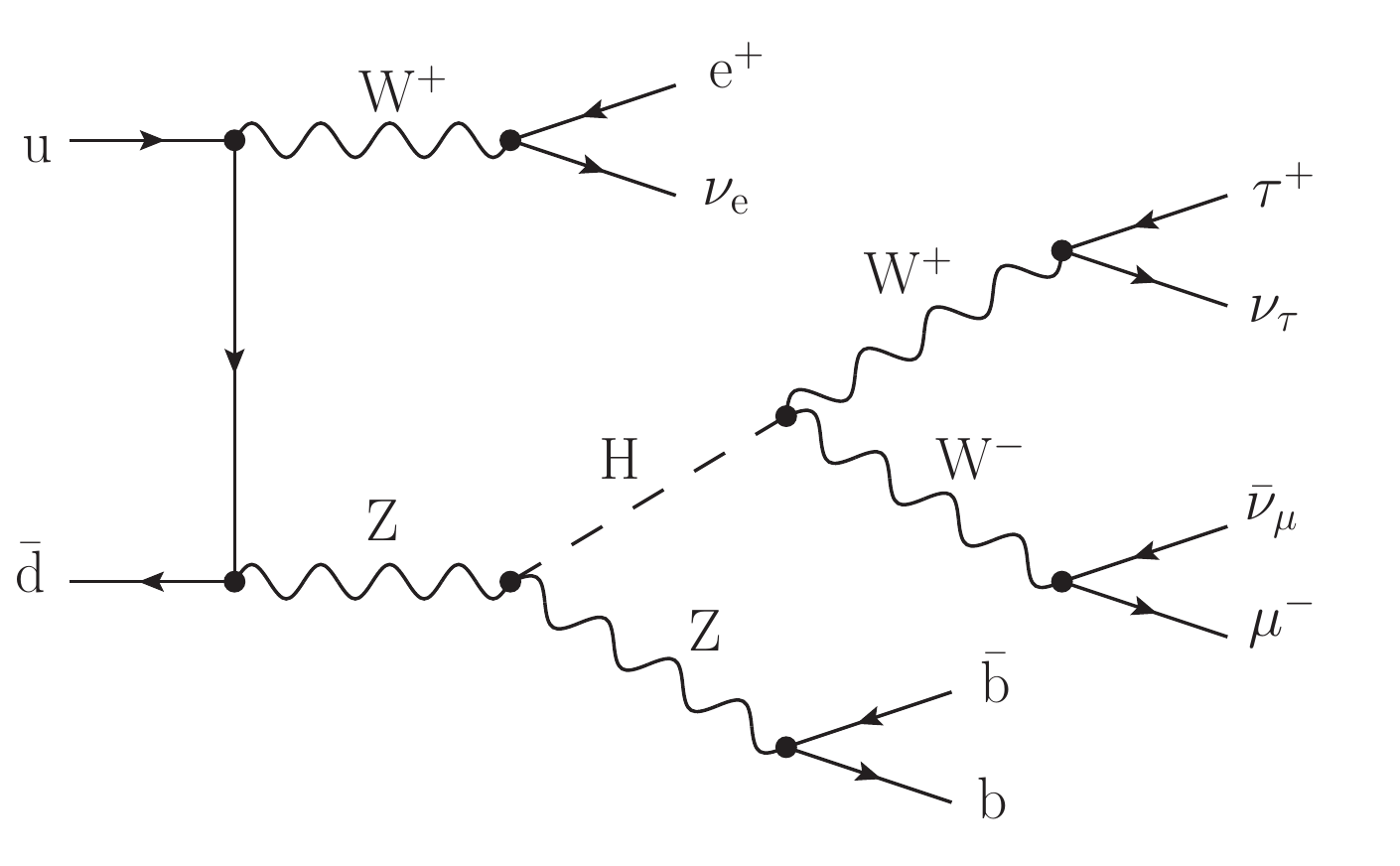}
  \caption{
    Sample diagrams contributing to $\loqcd$ (left) and to $\loew$ (right) cross-sections
    for off-shell $\Pt\overline{\Pt}\PW^+$ production in the three-charged-lepton decay channel.
  }\label{fig:lo}
\end{figure*}

We consider the process $\Pp\Pp\rightarrow \Pe^+\nu_{\Pe}\tau^+\nu_{\tau}\mu^-\bar{\nu}_\mu\,\Pb\,\bar{\Pb}\, + X$.
At LO exclusively  quark-induced partonic channels contribute, while
at NLO the gluon--quark and photon--quark channels open up. The gluon--gluon
channel only enters the calculation at NNLO QCD.
This process, embedding as dominant resonant structure a $\Pt\bar{\Pt}$ pair
in association with a $\PW^+$ boson, receives contributions from three different coupling orders
at LO, as can be observed in \reffi{fig:orders}: the largest contribution (labelled $\loqcd$) is of order $\mc O(\as^2\alpha^6)$ ,
while the $\mc O(\alpha^8)$ contribution (labelled $\loew$) is roughly 1\% of $\loqcd$.
Sample diagrams are shown in \reffi{fig:lo}. The interference among QCD and EW diagrams,
of order $\mc O(\as\alpha^7)$, vanishes owing to colour algebra (in the case of a diagonal CKM
matrix with unit entries). Both at $\loqcd$ and at $\loew$, diagrams with two, one, or zero resonant
top/antitop quarks are present in the off-shell calculation.

At NLO, four coupling orders contribute to $\Pt\bar{\Pt}\PW$ hadro-production (see \reffi{fig:orders}).
The pure QCD corrections to $\loqcd$ (labelled $\nloone$), of order $\mc O(\as^3\alpha^6)$ ,
are the dominant ones at NLO. The corresponding virtual corrections involve up to 7-point functions,
while the real corrections are challenging due to the high multiplicity of particle in the final state.
The $\nloone$ corrections to off-shell $\Pt\bar{\Pt}\PW$, calculated by two independent groups
\cite{Bevilacqua:2020pzy,Denner:2020hgg}, are strongly dependent on the (renormalization
and factorization) central-scale choice and range between 10\% and 20\% of the $\loqcd$ cross-section.

\begin{figure*} 
  \centering
  \includegraphics[scale=0.3]{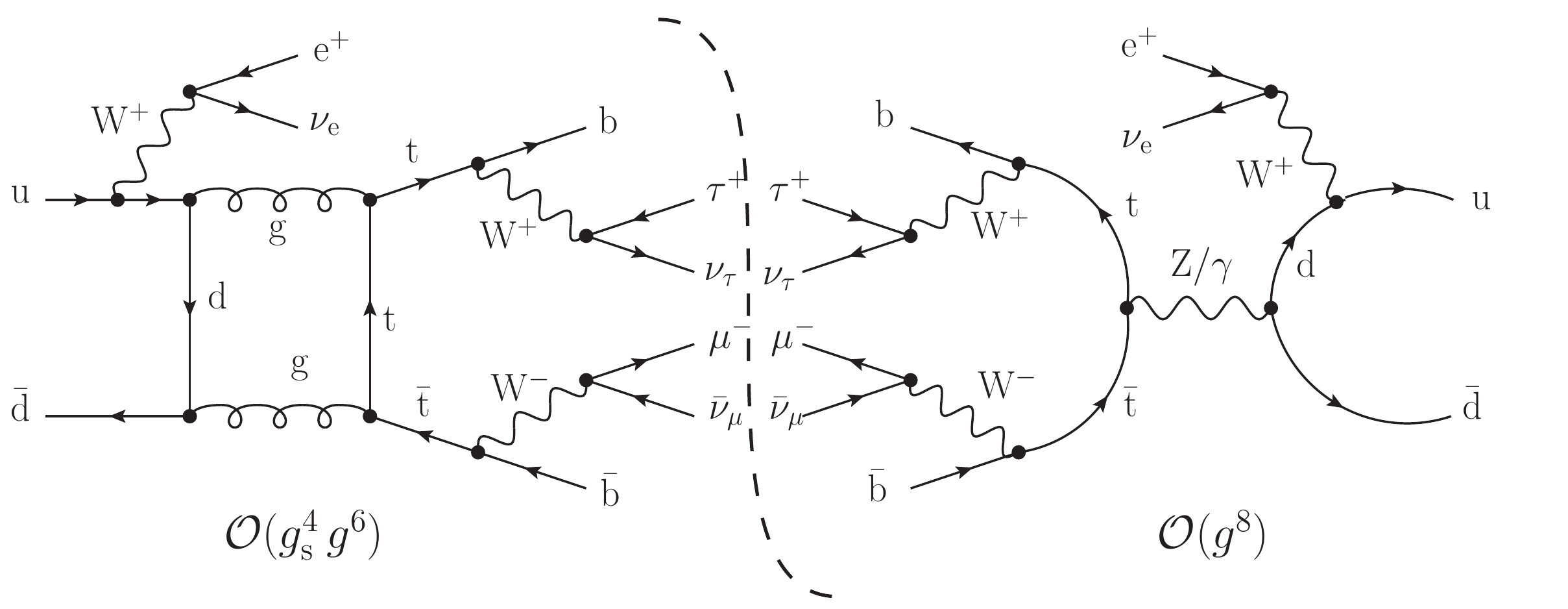}
  \includegraphics[scale=0.3]{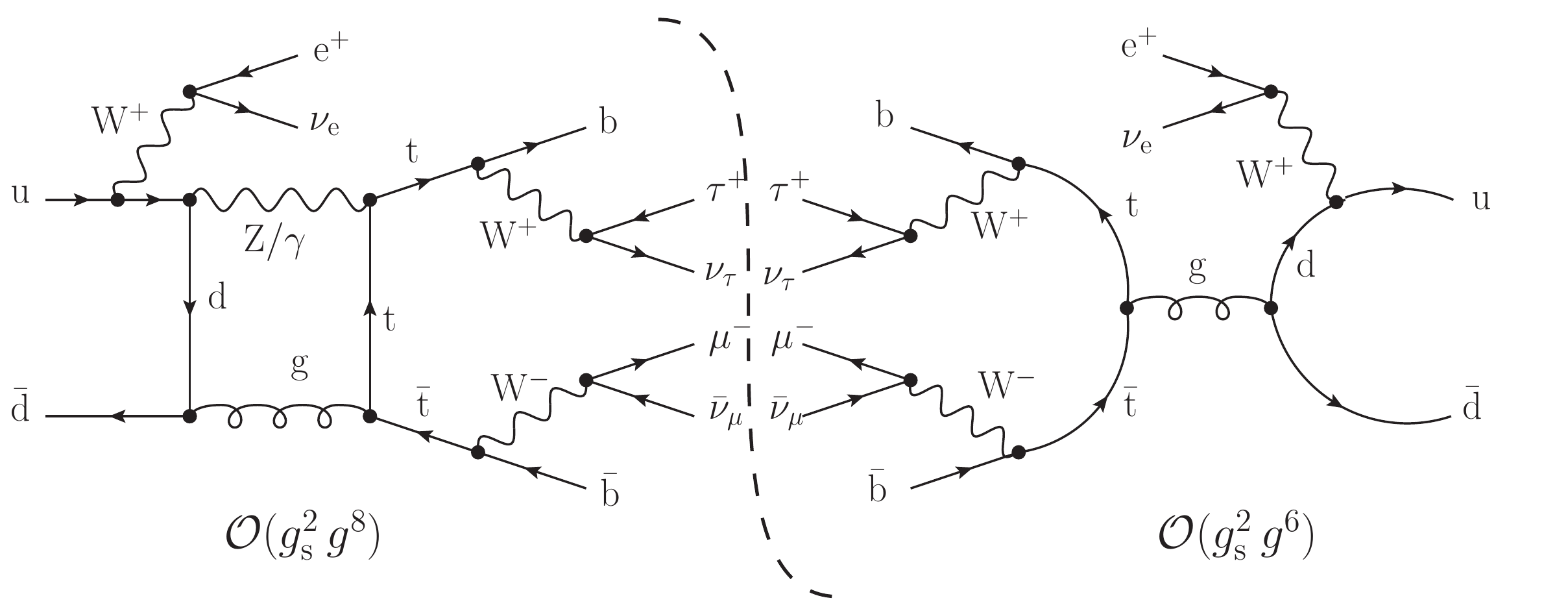}
  \caption{
    Sample $\mc O(\as^2\alpha^7)$ virtual corrections to
    off-shell $\Pt\overline{\Pt}\PW$ production
    in the three-charged-lepton decay channel: QCD corrections to the
    LO interference (left) and mixed contribution (right). }\label{fig:nlovirtclasses}
\end{figure*}

The corrections of order $\mc O(\as^2\alpha^7)$, labelled $\nlotwo$, come from the EW radiative corrections
to $\loqcd$ as well as from the QCD corrections to the LO interference, even though the contribution
of order $\mc O(\as\alpha^7)$ vanishes. The corresponding virtual corrections require up to 10-point functions
to be evaluated, and can be divided in two classes (see \reffi{fig:nlovirtclasses}): one-loop amplitudes
of order $\mc O(g_{\rm s}^4g^6)$ contracted with tree-level amplitudes of order $\mc O(g^8)$ and
one-loop amplitudes of order $\mc O(g_{\rm s}^2g^8)$ contracted with tree-level amplitudes of order
$\mc O(g_{\rm s}^2g^6)$. 
\begin{figure*}
  \centering
  \includegraphics[scale=0.28]{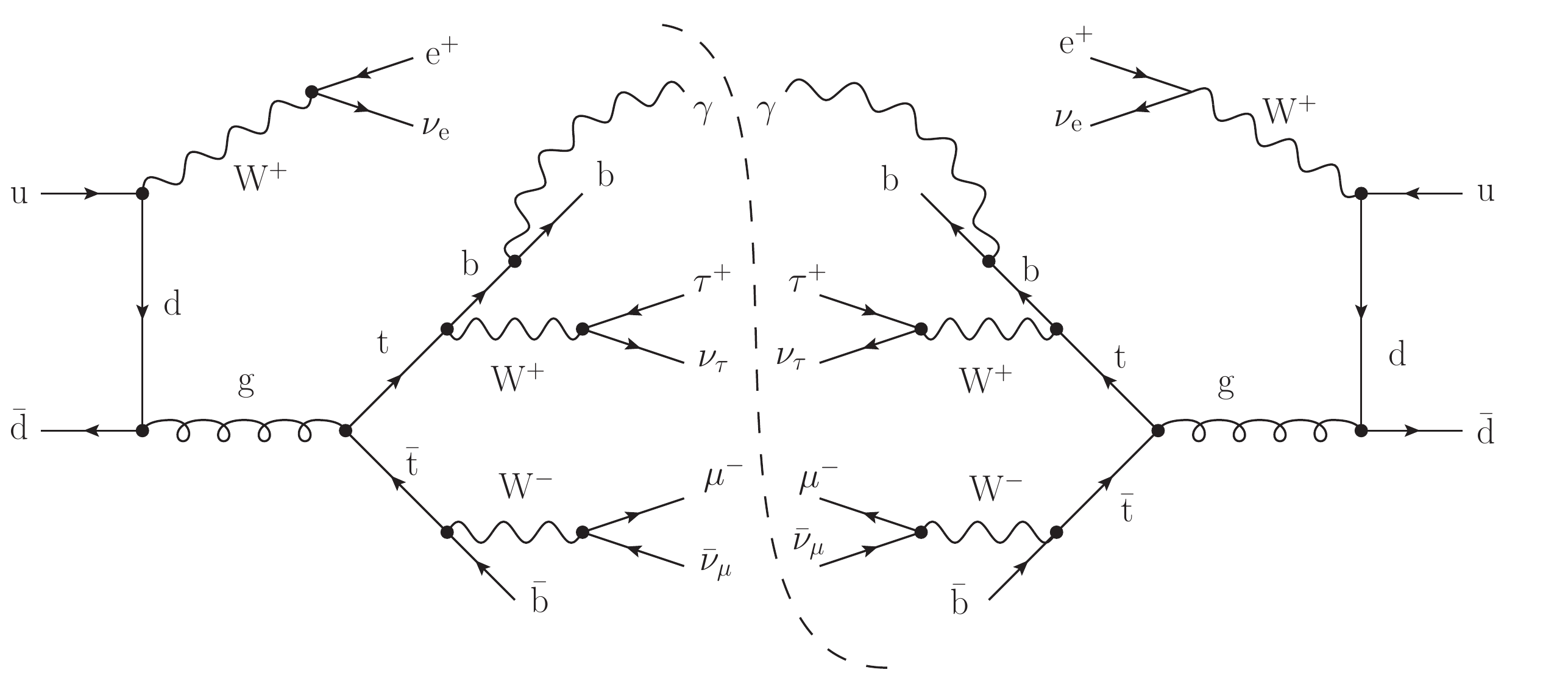}%
  \includegraphics[scale=0.28]{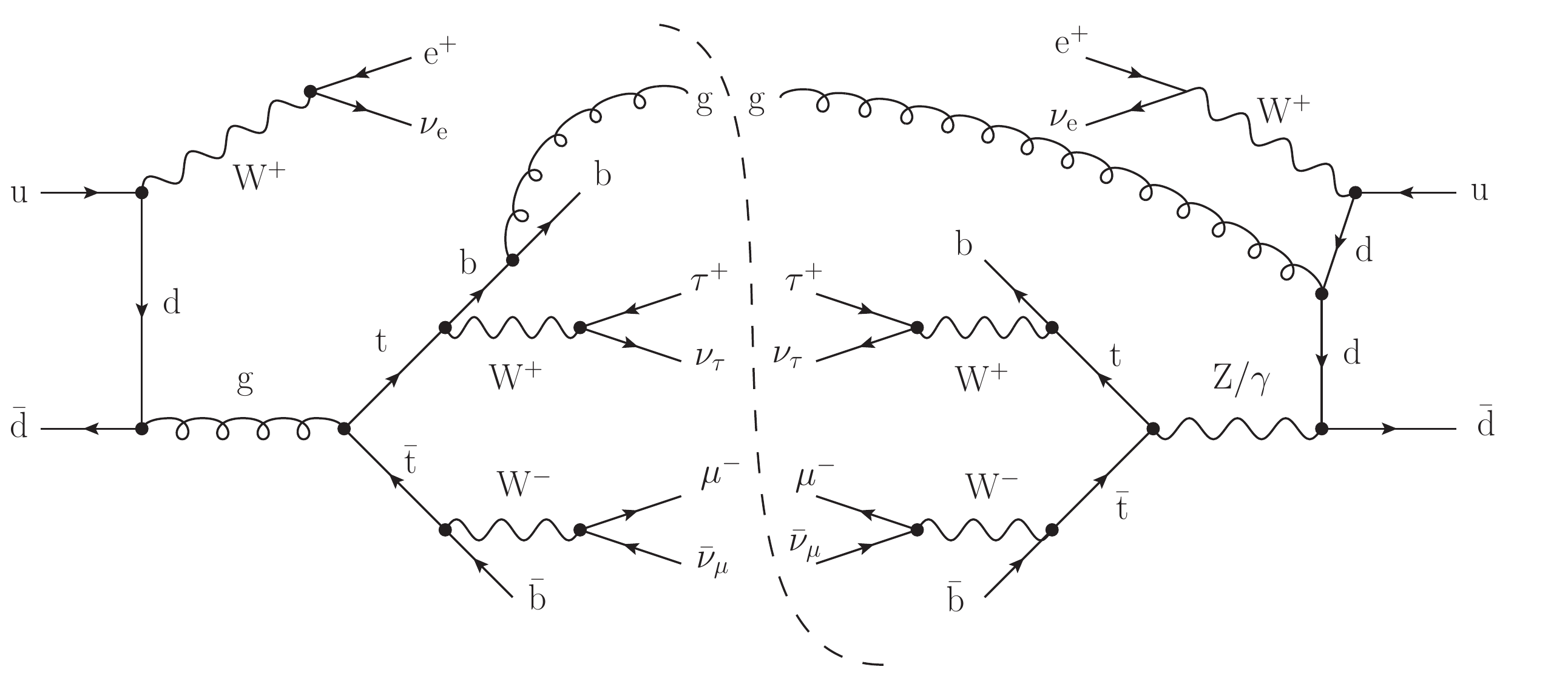}%
  \caption{
    Sample $\mc O(\as^2\alpha^7)$ real corrections to
    off-shell $\Pt\overline{\Pt}\PW$ production in the three-charged-lepton decay channel:
    EW corrections to $\loqcd$(left) and QCD corrections to the LO interference (right).
  }\label{fig:nlo2realinterf}
\end{figure*}
Analogously, the real corrections receive contributions from photon radiation off $\loqcd$
squared amplitudes as well as from gluon radiation off the LO interference. Sample diagrams
for real radiation at this perturbative order are shown in \reffi{fig:nlo2realinterf}.
It is crucial to include both EW corrections to $\loqcd$ and QCD corrections to the LO interference
to ensure the infrared (IR) finiteness of the NLO correction. In fact, the IR singularities
in one-loop amplitudes of order $\mc O(g_{\rm s}^2g^8)$ (right side of \reffi{fig:nlovirtclasses})
are cancelled by both classes of real-radiation contributions (\reffi{fig:nlo2realinterf}).

Since the LO interference vanishes, the corresponding EW corrections vanish as well. Therefore, the
$\nlothree$ corrections, of order $\mc O(\as\alpha^8)$, are pure QCD corrections to $\loew$.
Such corrections, although expected to be sub-leading w.r.t.\ the $\nloone$ and $\nlotwo$ ones
by $\as$-power counting arguments, give a larger contribution than the $\nlotwo$ ones
at the inclusive level \cite{Frixione:2015zaa,Frederix:2018nkq}, as they are dominated by real radiation contributions in the quark--gluon
partonic channel which embed the $\Pt\PW^+$ scattering as a sub-process \cite{Dror:2015nkp}.

The pure EW corrections to $\loew$, formally of order $\mc O(\alpha^9)$, have been shown to be
at the sub-per-mille level \cite{Frederix:2017wme,Frederix:2018nkq} and will be hardly relevant even at the
high-luminosity LHC. Therefore they are not considered in this context.

One-loop and tree-level amplitudes are calculated with \recola \cite{Actis:2012qn,Actis:2016mpe},
interfaced with \collier \cite{Denner:2016kdg} for the reduction and evaluation
of loop integrals. The Monte Carlo integration is performed via a multi-channel approach
with the \mocanlo code, which has already been utilized for several LHC processes with top quarks
\cite{Denner:2016jyo,Denner:2016wet} and now for $\Pt\overline{\Pt}\PW$ \cite{Denner:2020hgg,Denner:2021hqi}.
The subtraction of IR singularities is performed in the dipole scheme \cite{Catani:1996vz,Dittmaier:1999mb,Catani:2002hc}.
Top-quark and EW-boson masses, as well as the weak mixing angle are treated in the complex-mass scheme
\cite{Denner:1999gp,Denner:2000bj,Denner:2005fg,Denner:2006ic,Denner:2019vbn}.
Full off-shell matrix elements are considered at LO and NLO, including finite-width and non-resonant effects, as well as complete
spin correlations.
For more information on input parameters and details of the calculation, we refer to Sect.~(2.2) of \citere{Denner:2021hqi}.
    
\section{Phenomenological results}
We now present phenomenological results for a fiducial LHC setup that mimics the signal region defined
in a recent ATLAS measurement \cite{Aaboud:2019njj}.
For more details on the selection cuts we refer to Sect.~(2.3) of \citere{Denner:2021hqi}.
In Table~\ref{tab:integXS} we show integrated results for LO cross-sections and NLO corrections,
for three different choices of renormalization and factorization scale (labelled following the notation of \citere{Denner:2020hgg}),
\begin{align}\label{eq:scaledef}
&& \mu_0^{\rm (c)} = \frac13\left(\pt{\rm miss}+\sum_{i = {\Pb,\Pl}} \pt{i} \right)\,,\,\quad\mu_0^{\rm (d)} = \sqrt{\!\sqrt{\Mt^2+\pt{\Pt}^{\,2}}\,\sqrt{\Mt^2+\pt{\overline{\Pt}}^{\,2}}}\,,
\quad \mu_0^{\rm (e)} = \frac{\mu_0^{\rm (d)}}{2}\,,
\end{align}
where the ambiguity in identifying the top quark (for scale choices $\mu_0^{\rm (d)}$ and $\mu_0^{\rm (e)}$) is resolved by selecting decay products
that give an invariant mass closer to the top-quark pole mass.
\begin{table}
  \centering
  \small
\hspace*{-0.3cm}\begin{tabular}{c|cc|cc|cc}%
& \multicolumn{2}{c|}{$\mu_0^{\rm (c)}$ } & \multicolumn{2}{c|}{$\mu_0^{\rm (d)}$ } & \multicolumn{2}{c}{$\mu_0^{\rm (e)}$ } \\[0.5ex]
 \hline
 order  & $\sigma$ (fb)   &  ratio  & $\sigma$ (fb)   &  ratio  & $\sigma$ (fb)   &  ratio    \\[0.5ex] 
 \hline
  $\rm LO_{\rm QCD}$ ($\alpha_{\rm s}^2\alpha^6$) &     0.2218(1)$^{+25.3\%}_{-18.8\%}$  &    1  &   0.1948(1)$^{+23.9\%}_{-18.1\%}$  &   1 &   0.2414(1)$^{+26.2\%}_{-19.3\%}$  &   1  \\[0.5ex]          
  $\rm LO_{\rm EW}$ ($\alpha^8$)       &   0.002164(1)$^{+3.7\%}_{-3.6\%}$   &  0.010 & 0.002122(1)$^{+3.7\%}_{-3.6\%}$ & 0.011 & 0.002201(1)$^{+3.7\%}_{-3.6\%}$& 0.009  \\[0.5ex]
    \hline
    ${\rm NLO_1}$ ($\alpha_{\rm s}^3\alpha^6$)   &   0.0147(6)    &  0.066  & 0.0349(6) & 0.179 & 0.0009(7)& 0.004  \\[0.5ex] 
    ${\rm NLO_2}$ ($\alpha_{\rm s}^2\alpha^7$)&$\!\!\!\!-0.0122(3)$&$\!\!\!\!-0.055$ & $\!\!\!\!-0.0106(3)$& $\!\!\!\!-0.054$ & $\!\!\!\!-0.0134(4)$ & $\!\!\!\!-0.056$  \\[0.5ex]
    ${\rm NLO_3}$ ($\alpha_{\rm s}\alpha^8$)   &   0.0293(1)   &    0.131 & 0.0263(1)& 0.135 & 0.0320(1)&  0.133 \\[0.5ex]
    \hline
     $\rm LO_{\rm QCD}$+${\rm NLO_1}$ &   0.2365(6)$^{+2.9\%}_{-6.0\%}$ &   1.066&   0.2297(6)$^{+5.5\%}_{-7.3\%}$ &   1.179&   0.2423(7)$^{+3.5\%}_{-5.2\%}$ &   1.004\\[0.5ex]
   $\rm LO_{\rm QCD}$+${\rm NLO_2}$ & 0.2094(3)$^{+25.0\%}_{-18.7\%}$ &0.945 & 0.1840(3)$^{+23.8\%}_{-17.9\%}$& 0.946& 0.2277(4)$^{+25.9\%}_{-19.2\%}$  & 0.944\\[0.5ex]
   $\rm LO_{\rm EW}$+${\rm NLO_3}$ & 0.03142(4)$^{+22.2\%}_{-16.8\%}$ & 0.141 & 0.02843(6)$^{+20.5\%}_{-15.6\%}$ & 0.146& 0.03425(7)$^{+22.8\%}_{-17.0\%}$ &0.142\\[0.5ex]
  \hline
   LO+NLO &   0.2554(7)$^{+4.0\%}_{-6.5\%}$ &   1.151 &   0.2473(7)$^{+6.3\%}_{-7.6\%}$   &   1.270 &   0.2628(9)$^{+4.3\%}_{-5.9\%}$ &   1.089\\[0.5ex]
\end{tabular}
\caption{Fiducial LO cross-sections and NLO corrections (in fb) for three different central-scale choices [see Eq.~\eqref{eq:scaledef}]. Scale uncertainties from
7-point scale variations are shown in percentages. Ratios are relative to the $\loqcd$ cross-sections.}\label{tab:integXS}
\end{table}

The impact of $\nloone$ corrections relative to the $\loqcd$ cross-section is scale dependent,
and such corrections range between $+0.5\%$ and $+18\%$ depending on the central-scale choice.
At variance with $\nloone$, the relative $\nlotwo$ and $\nlothree$ corrections
are rather scale independent and amount to $-5\%$ and $+13\%$ of the $\loqcd$ result, respectively.
As expected from power counting, the $\loew$ cross-section is about 1\% of the $\loqcd$ one, while
the $\nlothree$ corrections are 10-times larger than $\loew$, giving more sizeable corrections
than the $\nlotwo$ ones, in spite of one power of $\as$ less. Also in the off-shell calculation,
hard real corrections embedding the $\Pt\PW$ scattering dominate the $\mc O(\as^3\alpha^6)$ perturbative order,
confirming the inclusive results \cite{Frixione:2015zaa,Frederix:2018nkq}. The NLO corrections,
obtained combining $\nloone$, $\nlotwo$, and $\nlothree$ in an additive way, range between $+9\%$
and $+27\%$, depending on the central-scale choice. The scale uncertainties at NLO are at the 5\% level and
are dominated by $\nloone$ corrections.
The $\nlotwo$ corrections have been calculated also in a very inclusive setup, obtaining similar
relative corrections to the $\loqcd$ cross-section ($-3\%$) as in on-shell calculations \cite{Frederix:2018nkq}.

\begin{figure*}
      \centering
      \includegraphics[scale=0.3]{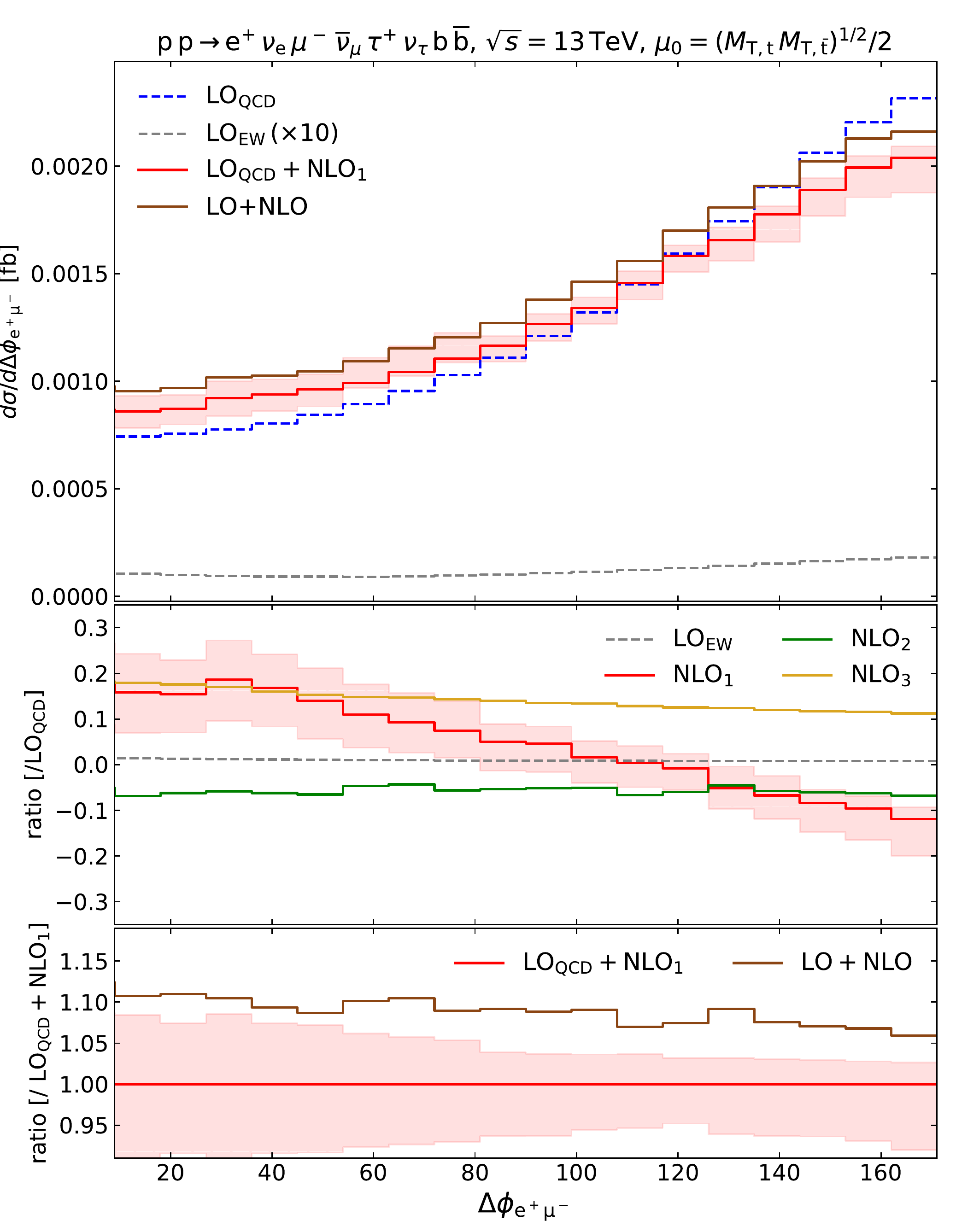}%
      \includegraphics[scale=0.3]{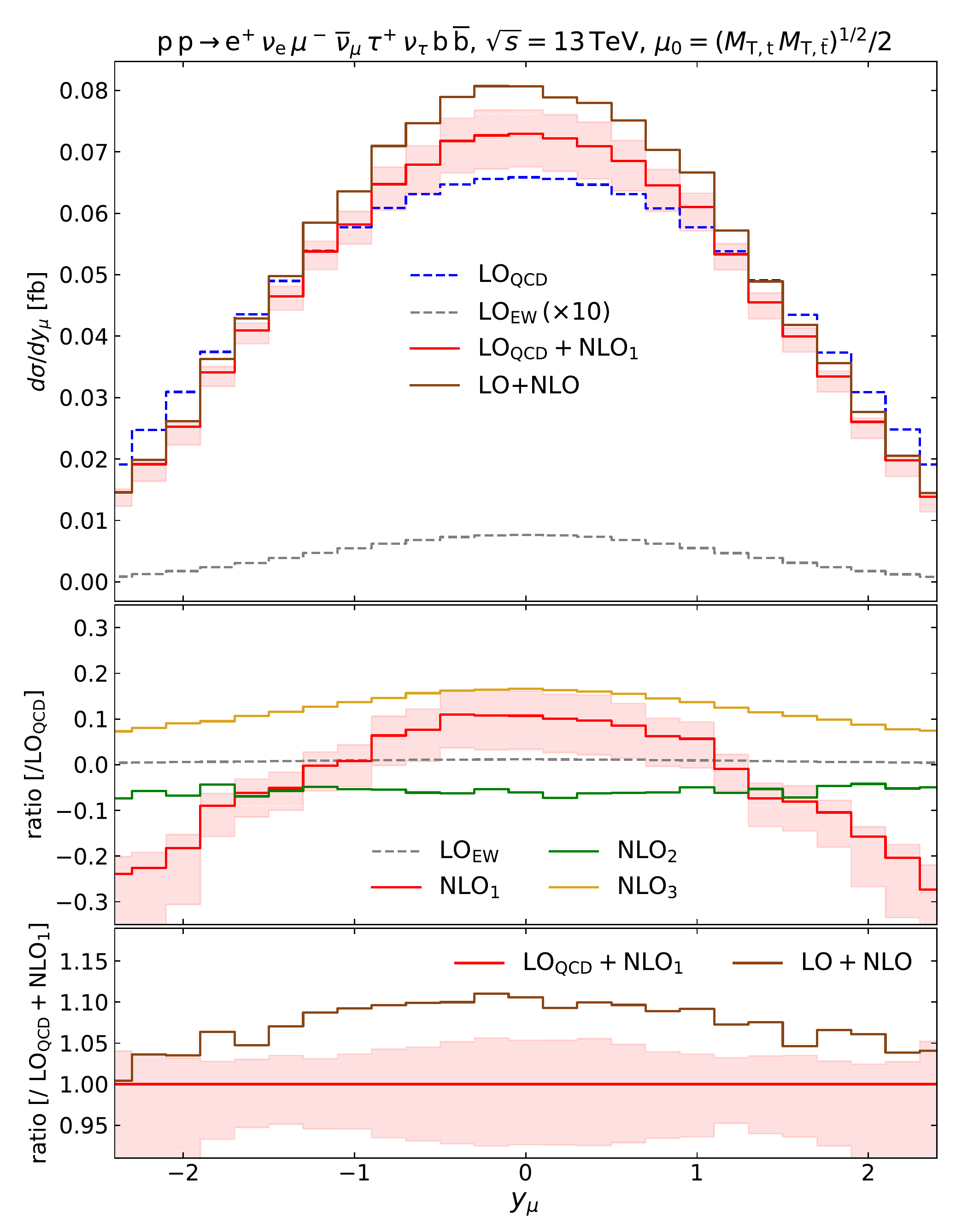}%
      \caption{Distributions in the azimuthal separation betwen the positron and the muon (left)
    and in the muon rapidity (right).
    Top: differential cross-sections in fb
    for $\rm LO_{\rm QCD}$, $\rm LO_{\rm EW}$, $\rm LO_{\rm QCD}+NLO_1$ and
    for complete NLO (sum of all LO cross-sections and NLO corrections).
    Middle: ratios of $\rm LO_{\rm EW}$, $\rm NLO_1$,
    $\rm NLO_2$, and $\rm NLO_3$ corrections over the $\rm LO_{\rm QCD}$ cross-section.
    Bottom: ratios of $\rm LO+NLO$ cross-section over the $\rm LO_{\rm QCD}+NLO_1$ one.
    Uncertainties from seven-point scale variations are shown for $\rm LO_{\rm QCD}+NLO_1$ distributions.}\label{fig:nloplots1}
    \end{figure*}

\begin{figure*}
  \centering
  \includegraphics[scale=0.3]{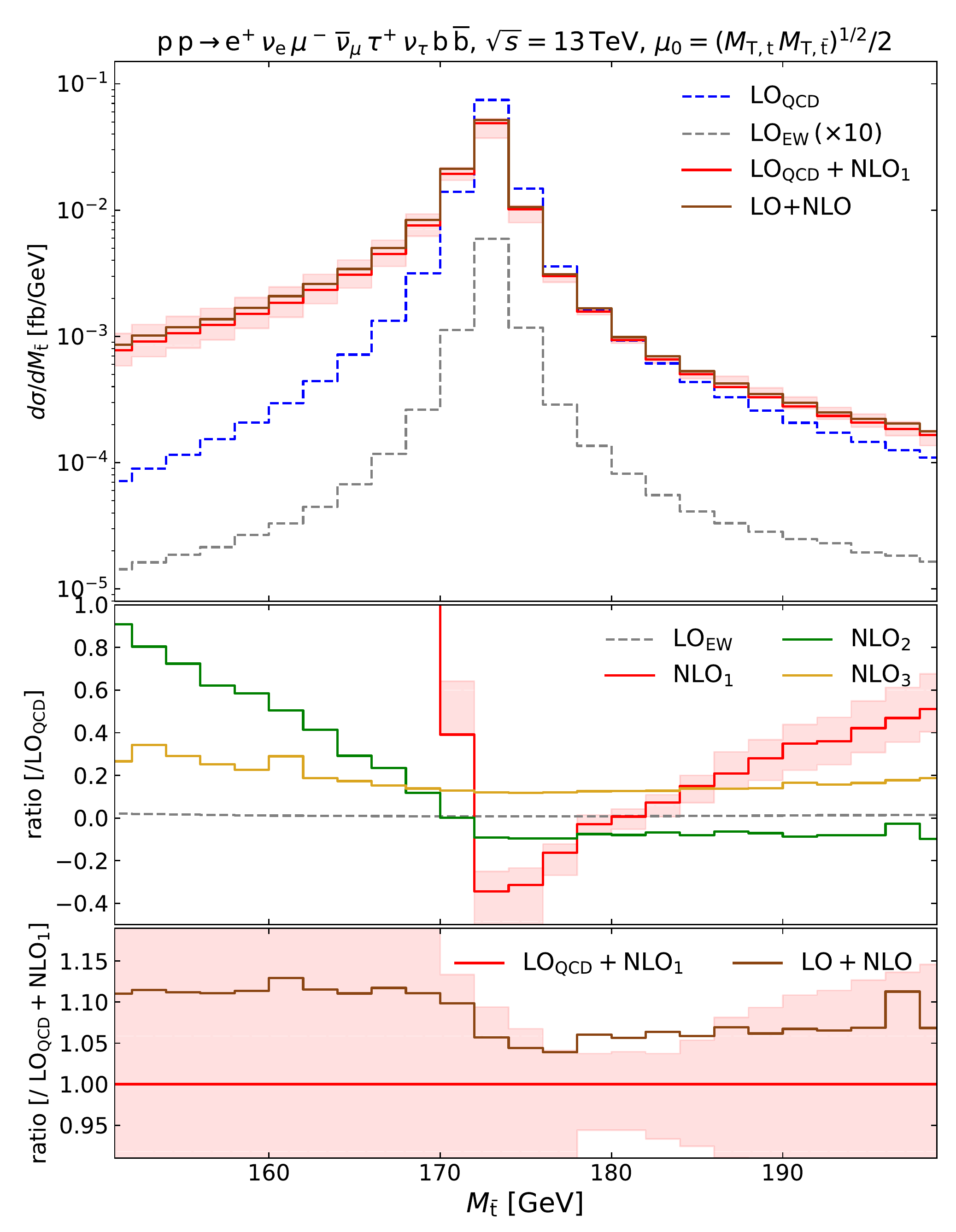}%
  \includegraphics[scale=0.3]{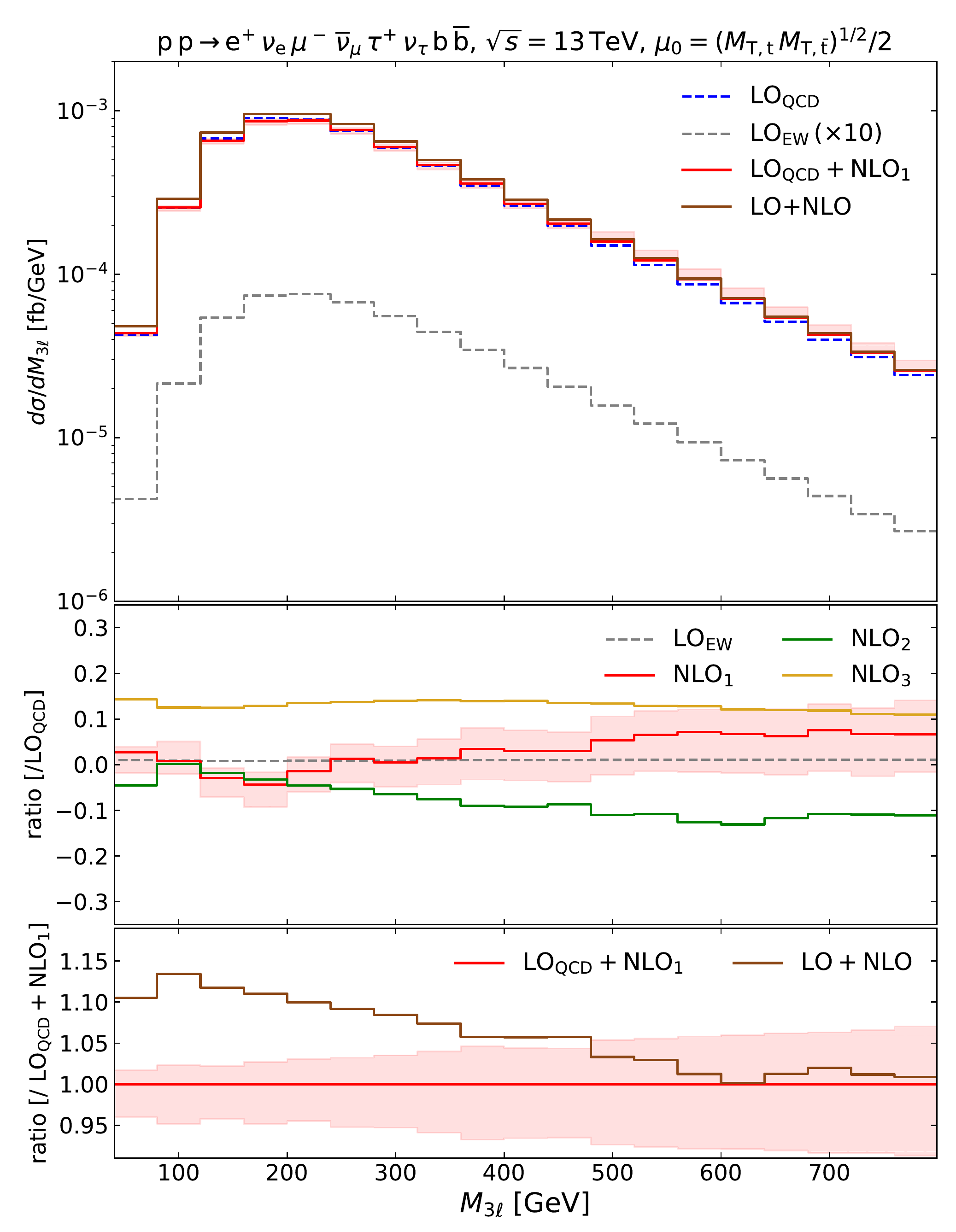}%
  \caption{ Distributions in the invariant mass of the antitop quark (left)
    and of the three-charged-lepton system (right). Same structure as \reffi{fig:nloplots1}. }\label{fig:nloplots2}
\end{figure*}

\begin{figure*}
      \centering
      \includegraphics[scale=0.3]{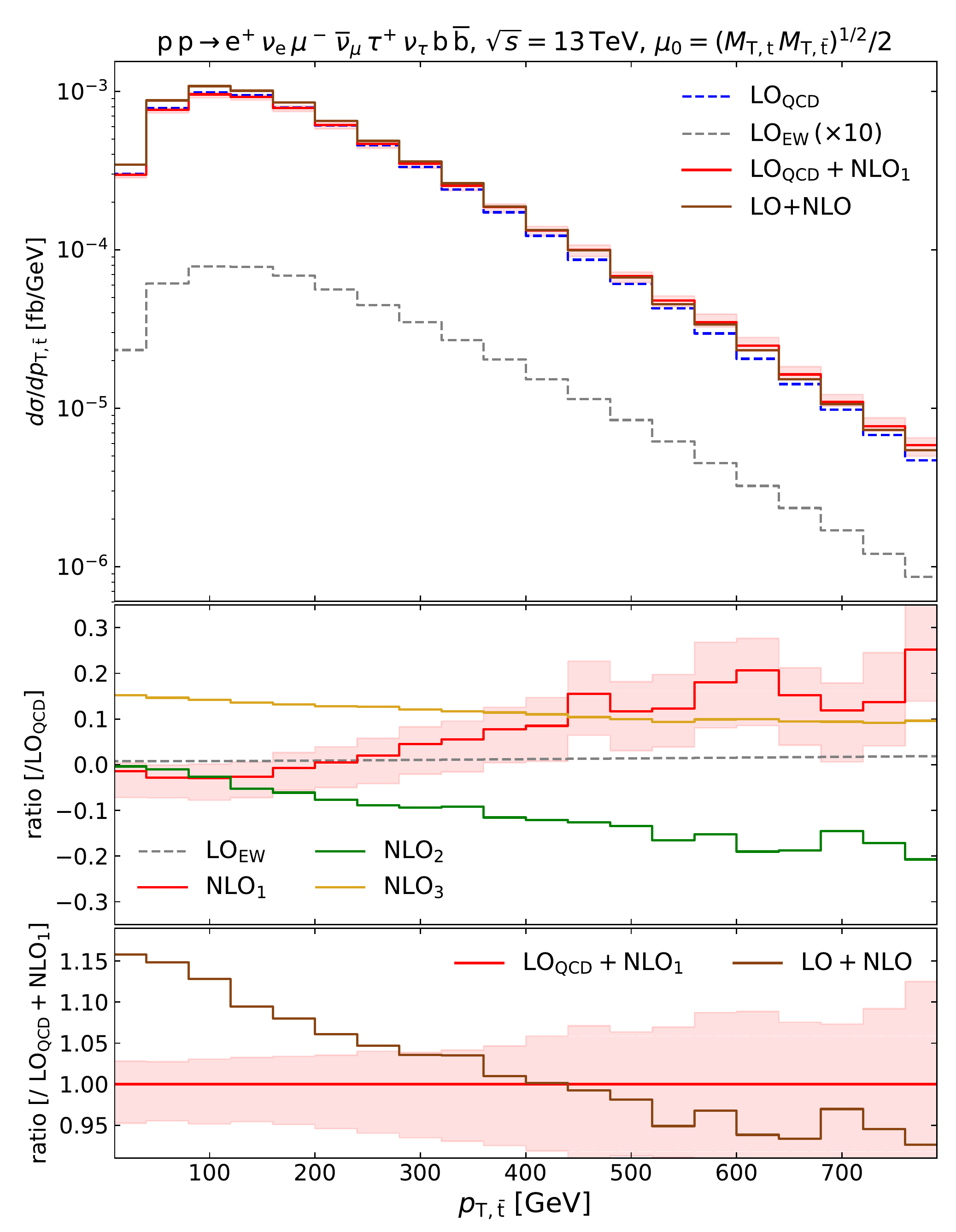}%
      \includegraphics[scale=0.3]{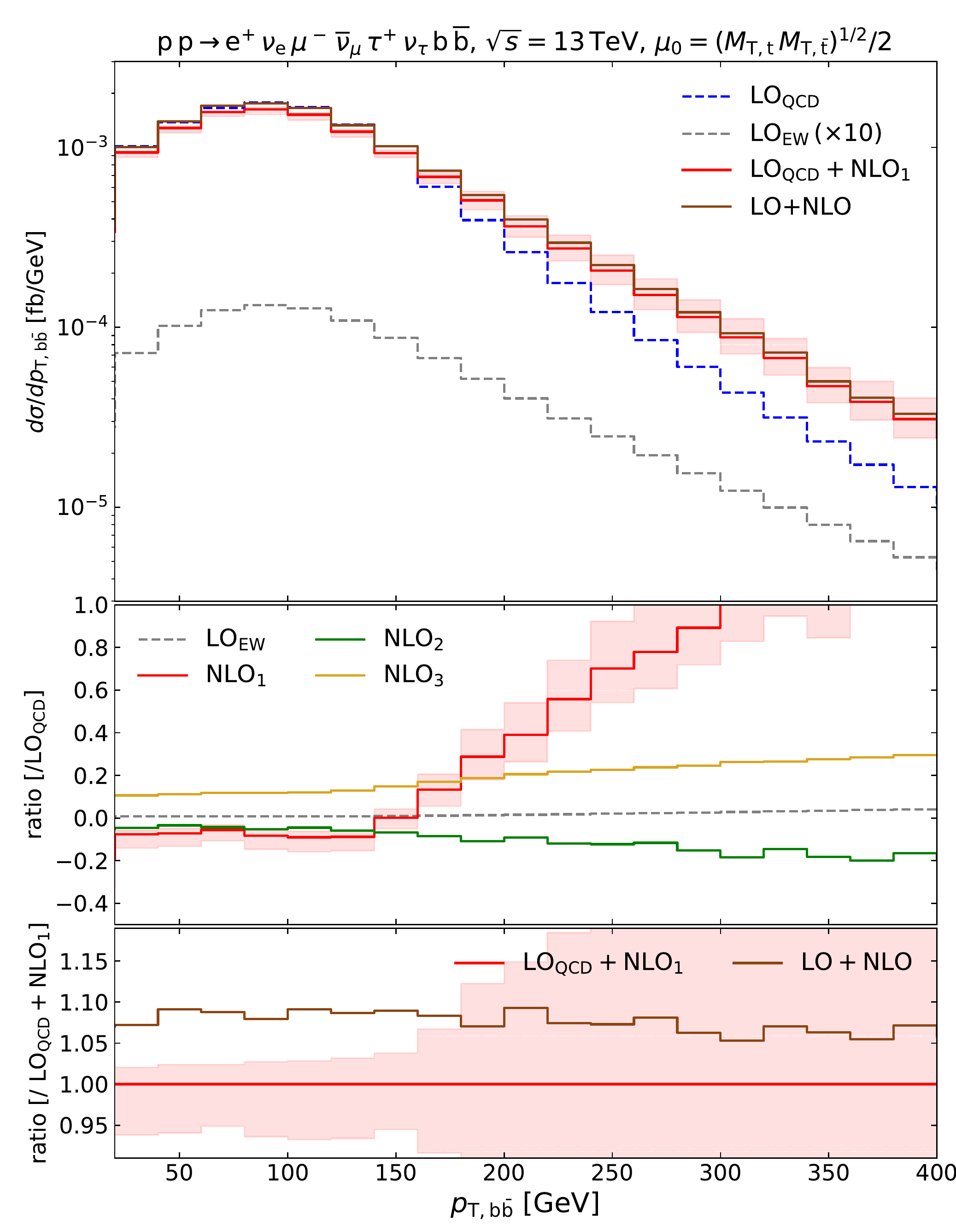}%
      \caption{ Distributions in the transverse momentum of the antitop quark (left)
    and of the two-b-jet system (right). Same structure as \reffi{fig:nloplots1}.}\label{fig:nloplots3}
    \end{figure*}

In \reffis{fig:nloplots1}--\ref{fig:nloplots3} we present differential results for the scale choice $\mu_0^{\rm (e)}$.
For exclusive observables, the interplay among the three NLO corrections can differ sizeably from
the results obtained for total cross-sections.

In the left panel of \reffi{fig:nloplots1} we show the distributions in the azimuthal separation between
the positron and the muon.
The NLO corrections increase the rate of events with small azimuthal separation.
The $\nloone$ corrections dinimish from $+18\%$ to $+11\%$ with constant slope over the distribution range,
while the $\nlotwo$ and $\nlothree$ ones give a rather constant shift to the
$\loqcd+\nloone$ cross-section.

In the right panel of \reffi{fig:nloplots1} we consider distributions in the muon rapidity.
This observable represents a good proxy for the rapidity of the antitop quark \cite{Denner:2021hqi}.
The muon is preferably produced with central rapidity. The $\nlotwo$ corrections give an almost flat
negative shift to the $\loqcd$ differential cross-section, while the relative $\nloone$ corrections feature a
variation of about 35\% in the rapidity range. The shape of $\nlothree$ corrections (relatively to $\loqcd$)
is similar to the one of $\nloone$ corrections. Such corrections give
a positive shift to the LO cross-section, ranging from $+16\%$ (central region) to $+8\%$ (forward regions).

In the left panel of \reffi{fig:nloplots2} the distributions in the antitop-quark invariant mass are shown.
The antitop-quark system ($\bar{\Pb}\mu^-\bar{\nu}_\mu$) is only known from the Monte Carlo truth, owing to the
presence of three neutrinos in the final state.
The $\nloone$ corrections are negative near the Breit--Wigner peak,
while they give a huge radiative enhancement to $\loqcd$ below the top-quark pole mass,
coming from real gluon radiation that is not clustered into b~jets.
A similar radiative tail, though less sizeable, is found also for $\nlotwo$ corrections.
For an invariant mass larger than the pole mass, $\nloone$ relative corrections increase towards
positive values while the $\nlotwo$ ones give an almost flat negative shift to $\loqcd$ ($-10\%$).
At variance with $\nloone$ and $\nlotwo$, the $\nlothree$ corrections are rather flat,
ranging between $+10\%$ at the peak and $+30\%$ in the tail region, due to the large quark--gluon
partonic channel, which features a light quark as final-state particle, which cannot result 
from the radiative decay of top/antitop quarks.

In the right panel of \reffi{fig:nloplots2} we consider the invariant mass of the system formed
by the three charged leptons. The QCD corrections ($\nloone$ and $\nlothree$) are rather flat,
while the $\nlotwo$ corrections, dominated by virtual EW corrections, negatively increase
towards large masses ($-10\%$ around $500\GeV$). Such a behaviour is driven by Sudakov logarithms, which become large at high
energy.

An analogous effect is found for transverse-momentum distributions, that are considered
in \reffi{fig:nloplots3}. 
In fact, large negative $\nlotwo$ corrections are found in the tail
of the transverse-momentum distribution for the reconstructed antitop quark ($-20\%$ at $800\GeV$)
as well as for the two-b-jet system ($-20\%$ at $350\GeV$).
The $\nlothree$ corrections are pretty flat for both observables considered in \reffi{fig:nloplots3},
ranging between $+10\%$ and $+30\%$ in the considered transverse-momentum ranges.
The $\nloone$ corrections increase by 25\% from small to large $p_{\rm T}$ of the antitop quark,
while they become much larger at moderate values of the transverse momentum of the two-b-jet system.
The two-b-jet system is correlated to the $\Pt\bar{\Pt}$ system \cite{Denner:2020hgg},
which recoils against a $\PW^+$ boson and thus receives large contributions from unclustered real QCD radiation.
The combined NLO corrections to the $\Pb\bar{\Pb}$-system transverse momentum is almost vanishing due to
cancellation among different contributions in the soft region of the spectrum, while it is dominated by $\nloone$ corrections
for $\pt{\Pb\Pb}>150\GeV$.

Both angular (\reffi{fig:nloplots1}) and energy-dependent distributions (\reffis{fig:nloplots2}--\ref{fig:nloplots3})
show that the combined NLO predictions exceed the scale uncertainties of the NLO QCD
results ($\loqcd+\nloone$) also in most populated kinematic regions,
\eg soft transverse momentum and central rapidity.


\section{Conclusion}
We have presented the first calculation including complete off-shell effects at
NLO QCD \sloppy [$\mc O(\as^3\alpha^6)$] and subleading NLO orders [$\mc O(\as^2\alpha^7)$ and $\mc O(\as\alpha^8)$]
for the hadronic production of $\Pt\bar{\Pt}\PW^+$ in the three-charged-lepton decay channel.
The $\nloone$ corrections range between $+0.5\%$ and $+18\%$, depending on the central-scale choice.
The scale uncertainties are reduced from
25\% to 5\% from LO to NLO, driven by the $\nloone$ corrections. The $\nlotwo$ corrections are negative
(about $-5\%$) and independent of the scale choice (relative to the LO cross-section), but they
become large (up to $-20\%$) for high-energy regimes in transverse-momentum and invariant-mass
distributions. The $\nlothree$ corrections are also large, ranging between $+10\%$ and $+30\%$
in the considered distributions, and are rather scale independent. These corrections are dominated
by gluon--quark-induced contributions embedding the $\Pt\PW$-scattering process.
The inclusion of $\nlotwo$ and $\nlothree$ corrections is necessary for the modelling of $\Pt\bar{\Pt}\PW^+$
production, as all NLO contributions change sizeably distribution shapes.
Futhermore, the off-shell effects become important in the tails of several distributions.

\section*{Acknowledgements}
This work is supported by the German Federal Ministry for
Education and Research (BMBF) under contract no.~05H18WWCA1.



\bibliography{ttv}

\begin{thebibliography}{10}
\providecommand{\url}[1]{\texttt{#1}}
\providecommand{\urlprefix}{URL }
\expandafter\ifx\csname urlstyle\endcsname\relax
  \providecommand{\doi}[1]{doi:\discretionary{}{}{}#1}\else
  \providecommand{\doi}{doi:\discretionary{}{}{}\begingroup
  \urlstyle{rm}\Url}\fi
\providecommand{\eprint}[2][]{\url{#2}}

\bibitem{Dror:2015nkp}
J.~A. Dror, M.~Farina, E.~Salvioni and J.~Serra,
\newblock \emph{{Strong tW Scattering at the LHC}},
\newblock JHEP \textbf{01}, 071 (2016),
\newblock \doi{10.1007/JHEP01(2016)071},
\newblock \eprint{1511.03674}.

\bibitem{Buckley:2015lku}
A.~Buckley, C.~Englert, J.~Ferrando, D.~J. Miller, L.~Moore, M.~Russell and
  C.~D. White,
\newblock \emph{{Constraining top quark effective theory in the LHC Run II
  era}},
\newblock JHEP \textbf{04}, 015 (2016),
\newblock \doi{10.1007/JHEP04(2016)015},
\newblock \eprint{1512.03360}.

\bibitem{Bylund:2016phk}
O.~Bessidskaia~Bylund, F.~Maltoni, I.~Tsinikos, E.~Vryonidou and C.~Zhang,
\newblock \emph{{Probing top quark neutral couplings in the Standard Model
  Effective Field Theory at NLO in QCD}},
\newblock JHEP \textbf{05}, 052 (2016),
\newblock \doi{10.1007/JHEP05(2016)052},
\newblock \eprint{1601.08193}.

\bibitem{Maltoni:2014zpa}
F.~Maltoni, M.~Mangano, I.~Tsinikos and M.~Zaro,
\newblock \emph{{Top-quark charge asymmetry and polarization in
  $t\overline{t}W^\pm$ production at the LHC}},
\newblock Phys. Lett. B \textbf{736}, 252 (2014),
\newblock \doi{10.1016/j.physletb.2014.07.033},
\newblock \eprint{1406.3262}.

\bibitem{Maltoni:2015ena}
F.~Maltoni, D.~Pagani and I.~Tsinikos,
\newblock \emph{{Associated production of a top-quark pair with vector bosons
  at NLO in QCD: impact on ${t}\overline{{t}}{H} $ searches at the LHC}},
\newblock JHEP \textbf{02}, 113 (2016),
\newblock \doi{10.1007/JHEP02(2016)113},
\newblock \eprint{1507.05640}.

\bibitem{Sirunyan:2017uzs}
A.~M. Sirunyan \emph{et~al.},
\newblock \emph{{Measurement of the cross section for top quark pair production
  in association with a W or Z boson in proton-proton collisions at $\sqrt{s}
  ={}$13\,TeV}},
\newblock JHEP \textbf{08}, 011 (2018),
\newblock \doi{10.1007/JHEP08(2018)011},
\newblock \eprint{1711.02547}.

\bibitem{Aaboud:2019njj}
M.~Aaboud \emph{et~al.},
\newblock \emph{{Measurement of the $t\bar{t}Z$ and $t\bar{t}W$ cross sections
  in proton-proton collisions at $\sqrt{s}={}$13\,TeV with the ATLAS
  detector}},
\newblock Phys. Rev. \textbf{D99}, 072009 (2019),
\newblock \doi{10.1103/PhysRevD.99.072009},
\newblock \eprint{1901.03584}.

\bibitem{ATLAS-CONF-2019-045}
\emph{{Analysis of $t\bar{t}H$ and $t\bar{t}W$ production in multilepton final
  states with the ATLAS detector}},
\newblock Tech. Rep. ATLAS-CONF-2019-045, CERN, Geneva (2019).

\bibitem{CMS-PAS-HIG-17-004}
\emph{{Search for Higgs boson production in association with top quarks in
  multilepton final states at $\sqrt{s}={}$13\,TeV}},
\newblock Tech. Rep. CMS-PAS-HIG-17-004, CERN, Geneva (2017).

\bibitem{Frixione:2015zaa}
S.~Frixione, V.~Hirschi, D.~Pagani, H.~S. Shao and M.~Zaro,
\newblock \emph{{Electroweak and QCD corrections to top-pair hadroproduction in
  association with heavy bosons}},
\newblock JHEP \textbf{06}, 184 (2015),
\newblock \doi{10.1007/JHEP06(2015)184},
\newblock \eprint{1504.03446}.

\bibitem{Frederix:2017wme}
R.~Frederix, D.~Pagani and M.~Zaro,
\newblock \emph{{Large NLO corrections in $t\bar{t}W^{\pm}$ and
  $t\bar{t}t\bar{t}$ hadroproduction from supposedly subleading EW
  contributions}},
\newblock JHEP \textbf{02}, 031 (2018),
\newblock \doi{10.1007/JHEP02(2018)031},
\newblock \eprint{1711.02116}.

\bibitem{Frederix:2018nkq}
R.~Frederix, S.~Frixione, V.~Hirschi, D.~Pagani, H.-S. Shao and M.~Zaro,
\newblock \emph{{The automation of next-to-leading order electroweak
  calculations}},
\newblock JHEP \textbf{07}, 185 (2018),
\newblock \doi{10.1007/JHEP07(2018)185},
\newblock \eprint{1804.10017}.

\bibitem{Li:2014ula}
H.~T. Li, C.~S. Li and S.~A. Li,
\newblock \emph{{Renormalization group improved predictions for $t\bar{t}W^\pm$
  production at hadron colliders}},
\newblock Phys. Rev. D \textbf{90}, 094009 (2014),
\newblock \doi{10.1103/PhysRevD.90.094009},
\newblock \eprint{1409.1460}.

\bibitem{Broggio:2016zgg}
A.~Broggio, A.~Ferroglia, G.~Ossola and B.~D. Pecjak,
\newblock \emph{{Associated production of a top pair and a W boson at
  next-to-next-to-leading logarithmic accuracy}},
\newblock JHEP \textbf{09}, 089 (2016),
\newblock \doi{10.1007/JHEP09(2016)089},
\newblock \eprint{1607.05303}.

\bibitem{Kulesza:2018tqz}
A.~Kulesza, L.~Motyka, D.~Schwartl{\"a}nder, T.~Stebel and V.~Theeuwes,
\newblock \emph{{Associated production of a top quark pair with a heavy
  electroweak gauge boson at NLO$+$NNLL accuracy}},
\newblock Eur. Phys. J. C \textbf{79}, 249 (2019),
\newblock \doi{10.1140/epjc/s10052-019-6746-z},
\newblock \eprint{1812.08622}.

\bibitem{Broggio:2019ewu}
A.~Broggio, A.~Ferroglia, R.~Frederix, D.~Pagani, B.~D. Pecjak and I.~Tsinikos,
\newblock \emph{{Top-quark pair hadroproduction in association with a heavy
  boson at NLO+NNLL including EW corrections}},
\newblock JHEP \textbf{08}, 039 (2019),
\newblock \doi{10.1007/JHEP08(2019)039},
\newblock \eprint{1907.04343}.

\bibitem{Kulesza:2020nfh}
A.~Kulesza, L.~Motyka, D.~Schwartl{\"a}nder, T.~Stebel and V.~Theeuwes,
\newblock \emph{{Associated top quark pair production with a heavy boson:
  differential cross sections at NLO+NNLL accuracy}},
\newblock Eur. Phys. J. C \textbf{80}, 428 (2020),
\newblock \doi{10.1140/epjc/s10052-020-7987-6},
\newblock \eprint{2001.03031}.

\bibitem{vonBuddenbrock:2020ter}
S.~von Buddenbrock, R.~Ruiz and B.~Mellado,
\newblock \emph{{Anatomy of inclusive $t\bar t W$ production at hadron
  colliders}},
\newblock Phys. Lett. B \textbf{811}, 135964 (2020),
\newblock \doi{10.1016/j.physletb.2020.135964},
\newblock \eprint{2009.00032}.

\bibitem{Frederix:2021agh}
R.~Frederix and I.~Tsinikos,
\newblock \emph{{On improving NLO merging for $t \bar t W$ production}}
  (2021),
\newblock \eprint{2108.07826}.

\bibitem{Campbell:2012dh}
J.~M. Campbell and R.~K. Ellis,
\newblock \emph{{$t \bar{t} W^{\pm}$ production and decay at NLO}},
\newblock JHEP \textbf{07}, 052 (2012),
\newblock \doi{10.1007/JHEP07(2012)052},
\newblock \eprint{1204.5678}.

\bibitem{Garzelli:2012bn}
M.~V. Garzelli, A.~Kardos, C.~G. Papadopoulos and Z.~Tr{\'o}cs{\'a}nyi,
\newblock \emph{{$t\bar{t}W^{\pm}$ and $t\bar{t}Z$ Hadroproduction at NLO
  accuracy in QCD with Parton Shower and Hadronization effects}},
\newblock JHEP \textbf{11}, 056 (2012),
\newblock \doi{10.1007/JHEP11(2012)056},
\newblock \eprint{1208.2665}.

\bibitem{Frederix:2020jzp}
R.~Frederix and I.~Tsinikos,
\newblock \emph{{Subleading EW corrections and spin-correlation effects in
  $t\bar{t}W$ multi-lepton signatures}},
\newblock Eur. Phys. J. C \textbf{80}, 803 (2020),
\newblock \doi{10.1140/epjc/s10052-020-8388-6},
\newblock \eprint{2004.09552}.

\bibitem{1843174}
F.~F. Cordero, M.~Kraus and L.~Reina,
\newblock \emph{{Top-quark pair production in association with a $W^\pm$ gauge
  boson in the POWHEG-BOX}},
\newblock Phys. Rev. D \textbf{103}(9), 094014 (2021),
\newblock \doi{10.1103/PhysRevD.103.094014},
\newblock \eprint{2101.11808}.

\bibitem{Bevilacqua:2020pzy}
G.~Bevilacqua, H.-Y. Bi, H.~B. Hartanto, M.~Kraus and M.~Worek,
\newblock \emph{{The simplest of them all: $t\bar{t} W^\pm$ at NLO accuracy in
  QCD}},
\newblock JHEP \textbf{08}, 043 (2020),
\newblock \doi{10.1007/JHEP08(2020)043},
\newblock \eprint{2005.09427}.

\bibitem{Denner:2020hgg}
A.~Denner and G.~Pelliccioli,
\newblock \emph{{NLO QCD corrections to off-shell
  $\text{t}\bar{\text{t}}\text{W}^+$ production at the LHC}},
\newblock JHEP \textbf{11}, 069 (2020),
\newblock \doi{10.1007/JHEP11(2020)069},
\newblock \eprint{2007.12089}.

\bibitem{Bevilacqua:2020srb}
G.~Bevilacqua, H.-Y. Bi, H.~B. Hartanto, M.~Kraus, J.~Nasufi and M.~Worek,
\newblock \emph{{NLO QCD corrections to off-shell ${t\bar{t}W^\pm}$ production
  at the LHC: Correlations and Asymmetries}},
\newblock Eur. Phys. J. C \textbf{81}, 675 (2021),
\newblock \doi{10.1140/epjc/s10052-021-09478-x},
\newblock \eprint{2012.01363}.

\bibitem{Bevilacqua:2021tzp}
G.~Bevilacqua, H.~Y. Bi, F.~F. Cordero, H.~B. Hartanto, M.~Kraus, J.~Nasufi,
  L.~Reina and M.~Worek,
\newblock \emph{{On the modeling uncertainties of $t\bar{t}W^\pm$ multi-lepton
  signatures}}  (2021),
\newblock \eprint{2109.15181}.

\bibitem{Denner:2021hqi}
A.~Denner and G.~Pelliccioli,
\newblock \emph{{Combined NLO EW and QCD corrections to off-shell $\text {t}
  \overline{\text {t}}\text {W} $ production at the LHC}},
\newblock Eur. Phys. J. C \textbf{81}(4), 354 (2021),
\newblock \doi{10.1140/epjc/s10052-021-09143-3},
\newblock \eprint{2102.03246}.

\bibitem{Actis:2012qn}
S.~Actis, A.~Denner, L.~Hofer, A.~Scharf and S.~Uccirati,
\newblock \emph{{Recursive generation of one-loop amplitudes in the Standard
  Model}},
\newblock JHEP \textbf{04}, 037 (2013),
\newblock \doi{10.1007/JHEP04(2013)037},
\newblock \eprint{1211.6316}.

\bibitem{Actis:2016mpe}
S.~Actis, A.~Denner, L.~Hofer, J.-N. Lang, A.~Scharf and S.~Uccirati,
\newblock \emph{{RECOLA: REcursive Computation of One-Loop Amplitudes}},
\newblock Comput. Phys. Commun. \textbf{214}, 140 (2017),
\newblock \doi{10.1016/j.cpc.2017.01.004},
\newblock \eprint{1605.01090}.

\bibitem{Denner:2016kdg}
A.~Denner, S.~Dittmaier and L.~Hofer,
\newblock \emph{{COLLIER: a fortran-based Complex One-Loop LIbrary in Extended
  Regularizations}},
\newblock Comput. Phys. Commun. \textbf{212}, 220 (2017),
\newblock \doi{10.1016/j.cpc.2016.10.013},
\newblock \eprint{1604.06792}.

\bibitem{Denner:2016jyo}
A.~Denner and M.~Pellen,
\newblock \emph{{NLO electroweak corrections to off-shell top-antitop
  production with leptonic decays at the LHC}},
\newblock JHEP \textbf{08}, 155 (2016),
\newblock \doi{10.1007/JHEP08(2016)155},
\newblock \eprint{1607.05571}.

\bibitem{Denner:2016wet}
A.~Denner, J.-N. Lang, M.~Pellen and S.~Uccirati,
\newblock \emph{{Higgs production in association with off-shell top-antitop
  pairs at NLO EW and QCD at the LHC}},
\newblock JHEP \textbf{02}, 053 (2017),
\newblock \doi{10.1007/JHEP02(2017)053},
\newblock \eprint{1612.07138}.

\bibitem{Catani:1996vz}
S.~Catani and M.~Seymour,
\newblock \emph{{A general algorithm for calculating jet cross-sections in NLO
  QCD}},
\newblock Nucl. Phys. B \textbf{485}, 291 (1997),
\newblock \doi{10.1016/S0550-3213(96)00589-5},
\newblock [Erratum: {Nucl. Phys. B} {\bf 510} (1998) 503--504],
\newblock \eprint{hep-ph/9605323}.

\bibitem{Dittmaier:1999mb}
S.~Dittmaier,
\newblock \emph{{A general approach to photon radiation off fermions}},
\newblock Nucl. Phys. B \textbf{565}, 69 (2000),
\newblock \doi{10.1016/S0550-3213(99)00563-5},
\newblock \eprint{hep-ph/9904440}.

\bibitem{Catani:2002hc}
S.~Catani, S.~Dittmaier, M.~H. Seymour and Z.~Tr{\'o}cs{\'a}nyi,
\newblock \emph{{The dipole formalism for next-to-leading order QCD
  calculations with massive partons}},
\newblock Nucl. Phys. B \textbf{627}, 189 (2002),
\newblock \doi{10.1016/S0550-3213(02)00098-6},
\newblock \eprint{hep-ph/0201036}.

\bibitem{Denner:1999gp}
A.~Denner, S.~Dittmaier, M.~Roth and D.~Wackeroth,
\newblock \emph{{Predictions for all processes $e^+ e^-\to
  4\,\mathrm{fermions}+\gamma$}},
\newblock Nucl. Phys. \textbf{B560}, 33 (1999),
\newblock \doi{10.1016/S0550-3213(99)00437-X},
\newblock \eprint{hep-ph/9904472}.

\bibitem{Denner:2000bj}
A.~Denner, S.~Dittmaier, M.~Roth and D.~Wackeroth,
\newblock \emph{{Electroweak radiative corrections to $e^+ e^-\to W W\to
  4\,$fermions in double pole approximation: The RACOONWW approach}},
\newblock Nucl. Phys. \textbf{B587}, 67 (2000),
\newblock \doi{10.1016/S0550-3213(00)00511-3},
\newblock \eprint{hep-ph/0006307}.

\bibitem{Denner:2005fg}
A.~Denner, S.~Dittmaier, M.~Roth and L.~H. Wieders,
\newblock \emph{{Electroweak corrections to charged-current $e^+ e^-\to 4\,$
  fermion processes: Technical details and further results}},
\newblock Nucl. Phys. \textbf{B724}, 247 (2005),
\newblock \doi{10.1016/j.nuclphysb.2011.09.001,
  10.1016/j.nuclphysb.2005.06.033},
\newblock [Erratum: {Nucl. Phys. B} {\bf 854} (2012) 504],
\newblock \eprint{hep-ph/0505042}.

\bibitem{Denner:2006ic}
A.~Denner and S.~Dittmaier,
\newblock \emph{{The complex-mass scheme for perturbative calculations with
  unstable particles}},
\newblock Nucl. Phys. B Proc. Suppl. \textbf{160}, 22 (2006),
\newblock \doi{10.1016/j.nuclphysbps.2006.09.025},
\newblock \eprint{hep-ph/0605312}.

\bibitem{Denner:2019vbn}
A.~Denner and S.~Dittmaier,
\newblock \emph{{Electroweak Radiative Corrections for Collider Physics}},
\newblock Phys. Rept. \textbf{864}, 1 (2020),
\newblock \doi{10.1016/j.physrep.2020.04.001},
\newblock \eprint{1912.06823}.

\end{thebibliography}

\nolinenumbers

\end{document}